\documentclass[runningheads]{llncs}
\usepackage{graphicx}
\usepackage[dvipsnames]{xcolor}
\usepackage{pifont}
\usepackage{amsfonts, yhmath}
\usepackage{cleveref}
\usepackage{graphicx}
\usepackage{booktabs}

\definecolor{mygreen}{RGB}{0, 150, 0}

\newcommand{\mypar}[1]{\noindent\textbf{#1.}}
\begin{document}
%
\title{TADM: Temporally-Aware Diffusion Model for Neurodegenerative Progression on Brain MRI}
%
\titlerunning{TADM: Temporally-Aware Diffusion Model}
%
\author{Mattia Litrico\inst{1} \and
Francesco Guarnera\inst{1} \and
Valerio Giuffirda\inst{2} \and
Daniele Ravì\inst{3} \and
Sebastiano Battiato\inst{1}}

\authorrunning{M. Litrico et al.}
%

\institute{Department of Mathematics and Computer Science,University of Catania, Catania, Italy\\
\email{mattia.litrico@phd.unict.it, francesco.guarnera@unict.it, sebastiano.battiato@unict.it }
\and
University of Nottingham, Nottingham, United Kingdom \email{valerio.giuffrida@nottingham.ac.uk} \and
University of Hertfordshire: Hatfield, United Kingdom \email{d.ravi@herts.ac.uk}}

\maketitle              

\begin{abstract}

Generating realistic images to accurately predict changes in the structure of brain MRI can be a crucial tool for clinicians. Such applications can help assess patients' outcomes and analyze how diseases progress at the individual level. However, existing methods developed for this task present some limitations. Some approaches attempt to model the distribution of MRI scans directly by conditioning the model on patients' ages, but they fail to explicitly capture the relationship between structural changes in the brain and time intervals, especially on age-unbalanced datasets. Other approaches simply rely on interpolation between scans, which limits their clinical application as they do not predict future MRIs. To address these challenges, we propose a Temporally-Aware Diffusion Model (TADM), which introduces a novel approach to accurately infer progression in brain MRIs. TADM learns the distribution of structural changes in terms of intensity differences between scans and combines the prediction of these changes with the initial baseline scans to generate future MRIs. Furthermore, during training, we propose to leverage a pre-trained Brain-Age Estimator (BAE) to refine the model's training process, enhancing its ability to produce accurate MRIs that match the expected age gap between baseline and generated scans. Our assessment, conducted on 634 subjects from the OASIS-3 dataset, uses similarity metrics and region sizes computed by comparing predicted and real follow-up scans on 3 relevant brain regions. TADM achieves large improvements over existing approaches, with an average decrease of 24\% in region size error and an improvement of 4\% in similarity metrics. These evaluations demonstrate the improvement of our model in mimicking temporal brain neurodegenerative progression compared to existing methods. We believe that our approach will significantly benefit clinical applications, such as predicting patient outcomes or improving treatments for patients.


\end{abstract}
%
%
%

\section{Introduction}
\label{sec:intro}
The capability to forecast structural changes in brain MRIs over time is critical in medical imaging. It facilitates early disease detection and enhances clinical intervention, particularly for severe conditions like Alzheimer's or Parkinson's diseases \cite{porsteinsson2021diagnosis}. While Alzheimer's disease (AD) diagnosis commonly relies on neuropsychological and behavioural assessments, imaging data significantly aid in identifying characteristic disease effects on the brain, even in its early stages \cite{jack2018nia}. Recent advancements in Artificial Intelligence (AI) have driven the development of sophisticated spatial-temporal disease progression models \cite{liu2006spatial}, empowering accurate prediction of brain structural progression. In particular, generative models have been proposed to simulate future MRI scans starting from past MRI scans used as inputs. One of the AI solutions employed in this context involves the training of Generative Adversarial Networks (GANs). For example, 4D-DANINet \cite{ravi2019degenerative} utilizes adversarial training alongside a series of biologically informed constraints to enhance the generation process. Another approach proposed in \cite{counterfactual_pred} uses a conditional 3D GAN with morphology constraints to predict deformations, instead of directly manipulating image pixels. More recently, approaches based on Denoising Diffusion Probabilistic Models (DDPMs) have demonstrated exceptional performance in this domain. For instance, in \cite{SADM}, a diffusion model is combined with a transformer network. The transformer generates a latent representation from a sequence of input MRIs, which is used to condition the generation process of the diffusion model. Another notable solution is Diffusion Deformable Model (DDM) \cite{DDM}, which introduces a methodology to combine a diffusion and a deformation module to generate images that interpolate between two MRI scans. Similarly, DiffuseMorph \cite{DiffuseMorph} trains a diffusion model to estimate a deformation field between two scans. This deformation field facilitates the translation of one input image into another, thereby enabling the generation of interpolated scans.

Due to the complexity of age-related changes in brain morphology during disease progression \cite{dickstein2007changes}, all these approaches often fail to accurately capture the corresponding temporal evolution, facing the following limitations: (i) approaches that operate conditioning on patient's age, such as \cite{ravi2019degenerative,counterfactual_pred}, do not explicitly capture the relationship between structural changes in brain MRIs and the time interval over which these changes occur; additionally, they require age-balanced datasets, which are often lacking in real-world applications; (ii) approaches based on interpolation, including DDM \cite{DDM} and DiffuseMorph \cite{DiffuseMorph}, have limitations in their capability to generate MRIs beyond the two input scans, reducing their relevance in clinical applications that require predicting future scans; and (iii) other approaches, like SADM \cite{SADM}, require longitudinal data at inference time, limiting again their application in real-world contexts where a series of scans for the patient are not available.

To address these issues, we propose TADM, a novel diffusion-based approach for brain progression modelling which operates directly on T1-weighted MRIs. Our model is trained to learn the distribution of brain changes within a specified time interval. To achieve this, we employ a three-fold strategy. Firstly, TADM learns to predict the intensity difference between baseline and follow-up MRIs. This avoids the need to generate entirely new scans, reduces the complexity of the problem, and mitigates generation errors. Secondly, we condition the model on the age gap between the input and output scans rather than directly on the output age, aiming to better learn the relationship between observed differences and the time interval. Given that the same age gap can arise between scans acquired at different ages, conditioning on age gap avoids the necessity of including samples from every age group in the training set. This is particularly beneficial when the dataset has limited samples in some age groups. Lastly, we propose to leverage a Brain-Age Estimator (BAE) to predict age differences between two scans. During training, these predicted age gaps are used in the loss function of our model, allowing the generation of images that accurately reflect the expected age gaps between the inputs and the predictions.

We evaluate our method on the OASIS-3 dataset \cite{lamontagne2019oasis}. Specifically, we compute similarity metrics and region size in three areas of the brain to estimate the difference between real and predicted follow-up brain images. TADM improves similarity metrics by 4\% and obtains the best performance on estimating region size reducing the error by 24\%. Additionally, our qualitative analysis shows visual improvements in our approach in terms of better mimicking the temporal progression of brains. 

In conclusion, the contributions of this work are: (i) introducing TADM, a diffusion-based approach for modelling brain progression trained on T1-weighted MRIs; (ii) learning the distribution of intensity differences between MRI scans to reduce the complexity of the generation process; (iii) conditioning on age gap to better capture the relationship between brain changes and time intervals; and (iv) proposing to leverage BAE to allow the generation of images that accurately reflect the expected age gap. The code will be available upon acceptance.


\section{Background}
\label{sec:rel_works}
Denoising Diffusion Probabilistic Models (DDPMs) \cite{ho2020denoising} are generative models that learn a Markov chain process to convert a Gaussian distribution into data distribution. During the diffusion process, Gaussian noise is added in successive steps to a sample $x_0$ from the given data distribution $q(x_0)$, to convert it into a latent variable distribution $q(x_t)$, as follows:

\begin{equation}
    q(x_{1},x_{2},\dots,x_{T}|x_{0}) = \prod_{t=1}^{T}q(x_{t}|x_{t-1})
\end{equation}
\begin{equation}
    q(x_{t}|x_{t-1})= \mathcal{N}(x_{t};\sqrt{1-\beta_{t}}x_{t-1},\beta_{t}\textbf{I})
\end{equation}
where $t \in {1, \ldots , T}$ is the diffusion step, $\mathcal{N}$ represents the Gaussian distribution, $\beta_{t}$ is a noise variance , and $\textbf{I}$ is the identity matrix.

Then, during the reverse process, the latent variable distribution $p_{\theta}(x_{t})$ is transformed progressively into the data distribution $p_{\theta}(x_{0})$, which is parameterized by $\theta$, by training the model to learn the following Gaussian transformations:
\begin{equation}
    p_{\theta}(x_{0},x_{1},\dots,x_{T-1}|x_{T}) = \prod_{t=1}^{T}p_{\theta}(x_{t-1}|x_{t})
\end{equation}
\begin{equation}
    p_{\theta}(x_{t-1}|x_{t})= \mathcal{N}(x_{t-1};\mu_{0}(x_{t},t),\sigma_{0}(x_{t},t)^{2}\textbf{I})
\end{equation}
\begin{equation}
    p(x_t)= \mathcal{N}(x_T; \textbf{0}, \textbf{I})
\end{equation}

\noindent where $\mu_{0}(x_{t},t)$ represents the mean of the Gaussian distribution, and $\sigma_{0}(x_{t},t)^{2}$ denotes the variance, at the $t$ reverse step.

\section{Proposed Method}
\label{sec:prop_meth}
\begin{figure}[t]
        \centering
        \includegraphics[width=\linewidth]{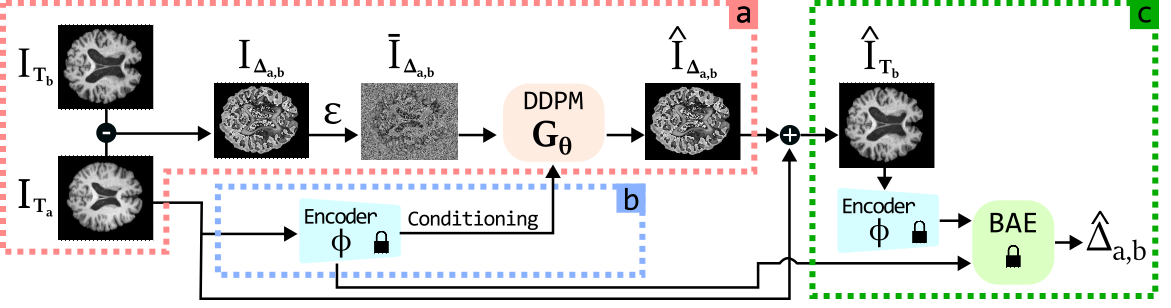}
    \caption{TADM comprises three main parts (surrounded by coloured dashed boxes). \textsc{\color{red}red:} The DDPM takes a residual image, calculated as the difference between two scans acquired at two the time points $T_a$ and $T_b$, on which random noise $\epsilon$ is applied. The DDPM is trained to denoise the residual image through several diffusion steps. The denoised residual image $\widehat{I}_{\Delta_{a,b}}$ and the scan $I_{T_a}$ are summed together to estimate the scan $\widehat{I}_{T_b}$ at time $T_{b}$. \textsc{\color{blue}blue:} Here, we encode the scan $I_{T_a}$ to extract a representation $z_a$ used to condition the DDPM, in conjunction with other patient-specific data. \textsc{\color{mygreen}green:} The estimated $\widehat{I}_{T_b}$ is provided to the encoder to extract the features $z_b$ that, together with the previously extracted features $z_a$, are provided as inputs to a BAE to predict the time interval $\widehat{\Delta}_{a,b}$. The \textsc{padlock} indicates a model with frozen parameters.}
    \label{fig:pipeline}
\end{figure}

In this section, we provide the details of TADM. Our pipeline is depicted in \Cref{fig:pipeline} and consists of the following blocks: (a) a DDPM, (b) an Encoder and (c) BAE.

During training, we use pairs of MRI scans denoted as $I_{T_a}$ and $I_{T_b}$, obtained from the same patient at two time points $T_a$ and $T_b$. These scans are used to compute a residual image $I_{\Delta_{a,b}} = I_{T_b} - I_{T_a}$ that we leverage to train the DDPM aimed at predicting residuals $\widehat{I}_{\Delta_{a,b}}$. To generated the output scan $\widehat{I}_{T_b}$ at time $T_b$ we then add the predicted residual $\widehat{I}_{\Delta_{a,b}}$ to the baseline scan $I_{T_a}$ at time $T_a$. Additionally, to achieve patient individualization, the DDPM is conditioned with a latent representation extracted by the encoder $\phi$ from $I_{T_a}$ and other patient-specific data (\textit{i.e.} cognitive status and age). Finally, we leverage BAE to predict the time interval $\Delta_{a,b} = T_b - T_a$ between $I_{T_a}$ and the estimated $\widehat{I}_{T_b}$ to aid the generation process of the DDPM. During inference, we use the scan $I_{T_a}$ at time $T_a$ together with the desired time interval $\Delta_{a,b'}$ as an input to generate a future unseen scan $I_{T_{b'}}$ at time $T_{b'}$ of the same patient.

\subsection{Conditioning the Diffusion Model}
\label{sec:conditioning}
We condition the DDPM to generate residual images using the following information: (i) the image representation $\phi(I_{T_a})$ extracted by the encoder $\phi$ on the baseline; (ii) the time interval $\Delta_{a,b}$; (iii) other patient's specific data.

\mypar{Image Representation} To obtain individualization at the subject level, we condition the model using a latent representation $z_a$ of the baseline scan $I_{T_a}$. In particular, the latent representation is obtained leveraging a pretrained encoder based on Residual-in-Residual Dense Blocks (RRDB) \cite{RRDB}.

\mypar{Time Interval (Age Gap)}
Conditioning the progression directly on age does not explicitly capture the relationship between structural changes in brain MRIs and the time interval over which these changes occur. Moreover, this strategy necessitates age-balanced datasets, which are difficult to observe in real-world scenarios. To tackle this limitation, we propose to condition the model using the age gap between scans $\Delta_{a,b}$.  Since the same age gap can occur between scans acquired at different ages, conditioning on the age gap eliminates the need for including samples from every age group in the training set. This is particularly advantageous when the dataset has limited samples in certain age groups. In our implementation, we encode $\Delta_{a,b}$ using positional encoding \cite{transformer} before incorporating it into the model.
 
\mypar{Patient-Specific Data} We also condition the model using the patient's cognitive status ($D$) and age at baseline ($A$). Indeed, age at baseline is crucial information as diseases progress at different rates over the course of ageing. However, by only using $\Delta_{a,b}$, our model would not capture such age-related progression information. Note that this is different from previous work on age conditioning, as they use age at the prediction rather than age at baseline.

\subsection{Leveraging BAE to improve the temporal awareness} \label{sec:tge}

To encourage the model to generate images that accurately reflect the expected age gap between the input and the prediction, we propose to leverage a BAE model \cite{jonsson2019brain} to predict the age gap between two MRI scans. BAE is trained offline on our training set and it is not further fine-tuned during the DDPM training. Specifically, given the baseline $I_{T_a}$ and the generated $\widehat{I}_{T_b}$ scans, the predicted age gap is computed as $\widehat{\Delta}_{a,b} = \Psi(\Phi(\widehat{I}_{T_b})) - \Psi(\Phi(I_{T_a}))$, where $\Psi$ is the BAE model. This information will be used later to train the DDPM and improve the generation. In particular, if the DDPM generates a scan $\widehat{I}_{T_b}$ that closely approximates the ground truth $I_{T_b}$, the predicted age gap $\widehat{\Delta}_{a,b}$ should closely match the actual age gap $\Delta_{a,b}$. Any error in the estimation of the age gap will be corrected through backpropagation in the diffusion model to refine the generation process.

\subsection{Overall Framework}
In this section, we will provide a complete overview of the framework.

\mypar{Training} During the diffusion process, the DDPM is trained to predict the noise $\epsilon$ added to the input $I_{\Delta_{a,b}}$. This is obtained by minimising the following objective:

\begin{equation}
    \label{eq:dml}
    \mathcal{L}^{DML} = \mathbb{E}_{\epsilon \sim \mathcal{N}(0,1), \mathbf{\bar{I}}_{\Delta_{a,b}}, t} \left[||G_\theta( \mathbf{\bar{I}}_{\Delta_{a,b}}, t; z_a, \Delta_{a,b}, A, D) - \epsilon ||^2_2 \right]
\end{equation}

\noindent where $G_\theta$ is the DDPM parametrized by $\theta$ and $t$ is the diffusion timestep.

Additionally, as mentioned in \Cref{sec:tge}, we incorporate the output from BAE as an additional term in the loss function of the DDPM. Specifically, we define the loss on the expected brain age gap as follows: 
\begin{equation}
    \label{eq:tge}
    \mathcal{L}^{BAE} = (\widehat{\Delta}_{a,b} - \Delta_{a,b})^2.
\end{equation}
Note that the gradient of this loss updates only the DDPM parameters $\theta$.

Finally, the overall loss is obtained by combining \cref{eq:dml,eq:tge} as follows:
\begin{equation}
    \mathcal{L}^{Tot} = \mathcal{L}^{DML} + \mathcal{L}^{BAE}
\end{equation}

\mypar{Inference} The model takes as input a baseline MRI $\mathbf{I}_{T_a}$ and predicts a follow-up MRI $\mathbf{I}_{T_{b'}}$ with an arbitrarily time interval $\Delta_{a,b'}$ with respect to the baseline. The reverse process starts from a Gaussian noise variable $X_T$ that is progressively denoised through $G_\theta( X_t, t; z_a, \Delta_{a,b'}, A, D)$.
The predicted residual image  $\mathbf{\hat{I}}_{\Delta_{a,b'}}$ is then added to the baseline MRI to generate the predicted follow-up. 



\section{Experimental Results}

\label{sec:exp_res}
\begin{table}[t]
    \caption{Comparison study: Results showing the performance in terms of image-base and region size in comparison to other methods.}
    \centering
    \setlength{\tabcolsep}{5pt}
    \def\arraystretch{1.5}
    \resizebox{\columnwidth}{!}{
    \begin{tabular}{l|cc|cccc}
    \toprule 
          &&& \multicolumn{4}{c}{Region Size Error (\%) $\downarrow$}\\
        Method  & SSIM $\uparrow$ & PSNR $\uparrow$ &Gray Matter & White Matter &Cerebrospinal Fluid & Total Brain   \\
        \midrule
        DiffuseMorph \cite{DiffuseMorph}  & 0.68 & 19.67 & 10.40 $\pm$ 3.45 & 3.49 $\pm$ 2.58 & 4.65 $\pm$ 2.80 & 46.30 $\pm$ 7.51 \\
        4D-DaniNet \cite{ravi2019degenerative}  & 0.65 & 16.99 & 2.21 $\pm$ 1.08 & 2.57 $\pm$ 1.98 & 3.12 $\pm$ 3.65 & 9.31 $\pm$ 8.72 \\ 
        DDM \cite{DDM}  & 0.69 & 19.59 & 2.44 $\pm$ 1.35 & 3.05 $\pm$ 2.74 & 4.37 $\pm$ 3.12 & 10.85 $\pm$ 11.64 \\
        
        \textbf{TADM (Proposed)}  & \textbf{0.72} & \textbf{20.51} & \textbf{1.69 $\pm$ 1.54}& \textbf{1.85 $\pm$ 2.20} & \textbf{2.70 $\pm$ 2.29} & \textbf{6.84 $\pm$ 5.00} \\
        \bottomrule
    \end{tabular}
    }
    
    \label{tab:oasis_results}
\end{table}

\mypar{Implementation Details} To evaluate our approach, we use 2,535 T1-weighted (T1w) brain MRIs from 634 subjects from the OASIS-3 dataset~\cite{lamontagne2019oasis}. Subjects are aged between $42$ and $95$ years and classified as cognitively normal, Mild Cognitive Impairment (MCI), and AD. We apply linear registration through the MNI152 template and skull removal using the FSL library~\cite{jenkinson2012fsl} to all the MRIs. The dataset is divided into training set (70\%), validation set (10\%) and test set (20\%). We used the validation set to optimize the hyperparameters of BAE.

Following \cite{li2022srdiff}, we adopt the U-Net as architecture of the diffusion model $G_\theta$ and the same hyperparameters.
Results of all baseline methods are obtained using their publicly available codes.

\mypar{Evaluation Metrics} We compute image-based metrics and region size in relevant brain areas to assess the performance of our method. Specifically, for the image-based metrics we use the Structural Similarity Index Measure (SSIM) and Peak Signal-to-Noise Ratio (PSNR) between the generated and the actual MRIs. On the other side, region sizes are used to evaluate the accuracy of disease progression. The regions considered in our experiment are: i) Gray Matter, ii) White Matter, and iii) Cerebrospinal Fluid (CSF). We use the FMRIB's Automated Segmentation Tool~\cite{jenkinson2012fsl} to compute the area of the regions which are expressed as percentages of the Total Brain to account for individual differences. The error is calculated as the mean absolute error between the area on the region from the predicted scan and ground truth MRIs.

\begin{figure}[t]
    \centering
    \includegraphics[width=1\linewidth]{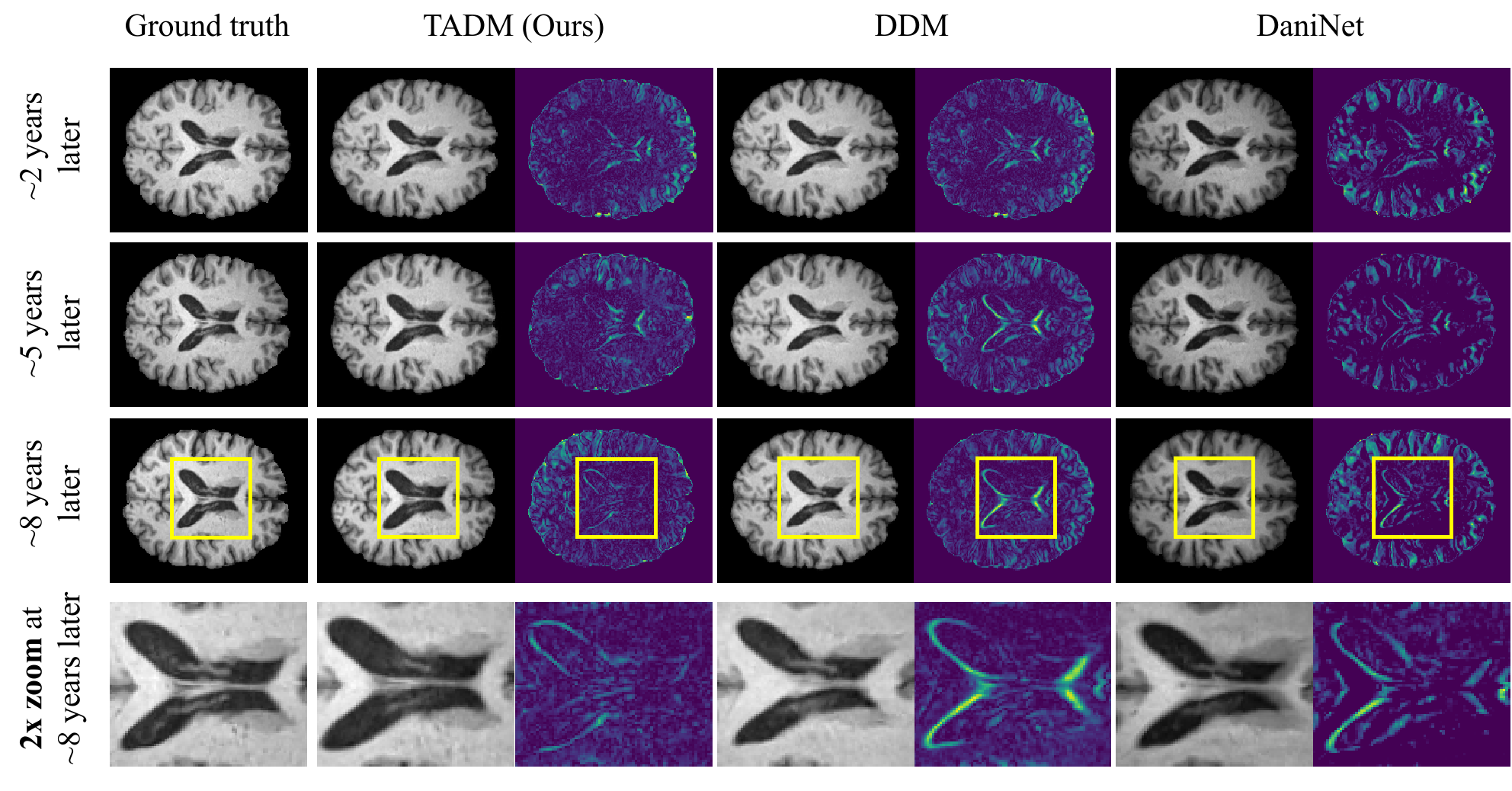}
    \caption{Comparison of the temporal progression on a 66-year-old subject with AD, obtained by our approach against other state-of-the-art methods. We show predicted slice-MRIs on the left and the corresponding error with the subject's real brain MRI on the right.}
    \label{fig:qualitative_results}
\end{figure}

\mypar{Comparison Results} In \Cref{tab:oasis_results}, we present quantitative results obtained by our method compared to other state-of-the-art approaches \cite{DiffuseMorph,DDM,ravi2019degenerative}. The results demonstrate that we outperform the state-of-the-art in SSIM and PSNR by $+0.03$ and $+0.84$, respectively. Results on the size of the region show that our method achieves the lowest error in all the considered brain regions. In particular, we reduce the error on grey and white matter regions by approximately $30\%$ and $8\%$, respectively. Regarding the CSF and Total Brain, the reduction of error is nearly $29\%$, demonstrating that our model generates high-quality follow-up scans compared to the current state-of-the-art. Additionally, in \Cref{fig:qualitative_results}, we show one example of qualitative results in predicting MRI scans at different time points. The figure shows that TADM offers a better approximation of the brain's temporal evolution compared to other methods. Specifically, our predictions depict a notably accurate alignment of the ventricular expansion over time. Finally, TADM exhibits fewer minor disparities in the brain cortex compared to other methods.

\begin{table}[t]
    \caption{Additional analysis: results showing the contribution of the different components of TADM, including BAE, the use of patient's specific data. Finally, we also show the impact of using the conditioning on age rather than age gap as we proposed.}
    \centering
    \setlength{\tabcolsep}{5pt}
    \def\arraystretch{1.5}
    \resizebox{\columnwidth}{!}{
    \begin{tabular}{l|cc|cccc}
        \toprule
        &&& \multicolumn{4}{c}{Region Size Error (\%) $\downarrow$}\\
        Method  & SSIM $\uparrow$ & PSNR $\uparrow$ & Gray Matter & White Matter & Cerebrospinal Fluid & Total Brain   \\
        \midrule
        TADM w/o patient's data & 0.71 & 20.32 & 1.78 $\pm$ 1.44 & 1.97 $\pm$ 2.14 & 2.72 $\pm$ 1.98 & 7.85 $\pm$ 5.17 \\
        TADM w/o BAE  & 0.69 & 20.08 & 2.44 $\pm$ 2.12 & 2.02 $\pm$ 2.13 & 3.85 $\pm$ 3.67 & 9.77 $\pm$ 8.23 \\
    
        TADM w/ age cond. & 0.68 & 19.71 & 4.12 $\pm$ 3.48 & 4.98 $\pm$ 2.45 & 5.65 $\pm$ 3.32 & 11.95 $\pm$ 7.34 \\
        \midrule
        \textbf{TADM} & \textbf{0.72} & \textbf{20.51} & \textbf{1.69 $\pm$ 1.54}& \textbf{1.85 $\pm$ 2.20} & \textbf{2.70 $\pm$ 2.29} & \textbf{6.84 $\pm$ 5.00} \\
        \bottomrule
    \end{tabular}
    }
    
    \label{tab:add_analysis}
\end{table}

\mypar{Additional analysis}

In this section, we explore the impact of integrating BAE into our pipeline and conditioning the model on patient's patient-specific data. Additionally, we evaluate the model's performance when it is conditioned on either the age or the age gap as proposed in our method. In the first row in \Cref{tab:add_analysis}, we show the results of our method without incorporating the patient's specific data. This outcome highlights that the absence of patient-specific data results in a minimal reduction in performance, indicating that the patient-specific data contributes minimally to the pipeline. When BAE is not used in our pipeline (second row in \Cref{tab:add_analysis}), we notice an evident decrease in performance. This indicates that leveraging BAE is essential to support the generation process. Lastly, in the third row, we observe that conditioning the model on age rather than age gap drastically reduces performance by an average of approximately $60\%$, demonstrating the effectiveness of our idea of conditioning on age gaps.


\section{Conclusion and Future Works}
\label{sec:conclusion}
In this paper, we propose TADM, a novel approach designed to accurately mimic brain neurodegenerative progression in MRIs. We evaluated TADM on the OASIS-3 dataset, demonstrating superior performance compared to existing approaches. TADM focuses on 2D scans, as we strive to develop a new data-driven pipeline capable of improving the accuracy of current methods. Nonetheless, our method can be easily extended to 3D scans.
Furthermore, our pipeline presents exciting prospects for data augmentation of underrepresented samples in medical imaging datasets. This feature holds significant promise, especially in the context of less common and more expensive modalities, \textit{e.g.} PET and CT scans, where the generation of synthetic samples remains a critical challenge.

%
%
%
\newpage
 \bibliographystyle{splncs04}
 \bibliography{main}

\begin{thebibliography}{10}
\providecommand{\url}[1]{\texttt{#1}}
\providecommand{\urlprefix}{URL }
\providecommand{\doi}[1]{https://doi.org/#1}

\bibitem{dickstein2007changes}
Dickstein, D.L., Kabaso, D., Rocher, A.B., Luebke, J.I., Wearne, S.L., Hof, P.R.: Changes in the structural complexity of the aged brain. Aging cell  \textbf{6}(3),  275--284 (2007)

\bibitem{ho2020denoising}
Ho, J., Jain, A., Abbeel, P.: Denoising diffusion probabilistic models. Advances in neural information processing systems  \textbf{33},  6840--6851 (2020)

\bibitem{jack2018nia}
Jack~Jr, C.R., Bennett, D.A., Blennow, K., Carrillo, M.C., Dunn, B., Haeberlein, S.B., Holtzman, D.M., Jagust, W., Jessen, F., Karlawish, J., et~al.: Nia-aa research framework: toward a biological definition of alzheimer's disease. Alzheimer's \& Dementia  \textbf{14}(4),  535--562 (2018)

\bibitem{jenkinson2012fsl}
Jenkinson, M., Beckmann, C.F., Behrens, T.E., Woolrich, M.W., Smith, S.M.: {FSL}. Neuroimage  \textbf{62}(2),  782--790 (2012)

\bibitem{jonsson2019brain}
J{\'o}nsson, B.A., Bjornsdottir, G., Thorgeirsson, T., Ellingsen, L.M., Walters, G.B., Gudbjartsson, D., Stefansson, H., Stefansson, K., Ulfarsson, M.: Brain age prediction using deep learning uncovers associated sequence variants. Nature communications  \textbf{10}(1), ~5409 (2019)

\bibitem{DiffuseMorph}
Kim, B., Han, I., Ye, J.C.: Diffusemorph: Unsupervised deformable image registration using diffusion model. In: Computer Vision – ECCV 2022: 17th European Conference, Tel Aviv, Israel, October 23–27, 2022, Proceedings, Part XXXI. p. 347–364. Springer-Verlag, Berlin, Heidelberg (2022)

\bibitem{DDM}
Kim, B., Ye, J.C.: Diffusion deformable model for 4d temporal medical image generation. In: Medical Image Computing and Computer Assisted Intervention – MICCAI 2022: 25th International Conference, Singapore, September 18–22, 2022, Proceedings, Part I. p. 539–548. Springer-Verlag, Berlin, Heidelberg (2022)

\bibitem{lamontagne2019oasis}
LaMontagne, P.J., Benzinger, T.L., Morris, J.C., Keefe, S., Hornbeck, R., Xiong, C., Grant, E., Hassenstab, J., Moulder, K., Vlassenko, A.G., et~al.: Oasis-3: longitudinal neuroimaging, clinical, and cognitive dataset for normal aging and alzheimer disease. MedRxiv pp. 2019--12 (2019)

\bibitem{li2022srdiff}
Li, H., Yang, Y., Chang, M., Chen, S., Feng, H., Xu, Z., Li, Q., Chen, Y.: Srdiff: Single image super-resolution with diffusion probabilistic models. Neurocomputing  \textbf{479},  47--59 (2022)

\bibitem{liu2006spatial}
Liu, D., Kelly, M., Gong, P.: A spatial--temporal approach to monitoring forest disease spread using multi-temporal high spatial resolution imagery. Remote sensing of environment  \textbf{101}(2),  167--180 (2006)

\bibitem{counterfactual_pred}
Pombo, G., Gray, R., Cardoso, M.J., Ourselin, S., Rees, G., Ashburner, J., Nachev, P.: Equitable modelling of brain imaging by counterfactual augmentation with morphologically constrained 3d deep generative models. Medical Image Analysis  \textbf{84},  102723 (2023). \doi{https://doi.org/10.1016/j.media.2022.102723}, \url{https://www.sciencedirect.com/science/article/pii/S1361841522003516}

\bibitem{porsteinsson2021diagnosis}
Porsteinsson, A., Isaacson, R., Knox, S., Sabbagh, M., Rubino, I.: Diagnosis of early alzheimer’s disease: clinical practice in 2021. The journal of prevention of Alzheimer's disease  \textbf{8},  371--386 (2021)

\bibitem{ravi2019degenerative}
Ravi, D., Alexander, D.C., Oxtoby, N.P., Initiative, A.D.N.: Degenerative adversarial neuroimage nets: generating images that mimic disease progression. In: International Conference on Medical Image Computing and Computer-Assisted Intervention. pp. 164--172. Springer (2019)

\bibitem{transformer}
Vaswani, A., Shazeer, N., Parmar, N., Uszkoreit, J., Jones, L., Gomez, A.N., Kaiser, L.u., Polosukhin, I.: Attention is all you need. In: Guyon, I., Luxburg, U.V., Bengio, S., Wallach, H., Fergus, R., Vishwanathan, S., Garnett, R. (eds.) Advances in Neural Information Processing Systems. vol.~30. Curran Associates, Inc. (2017)

\bibitem{RRDB}
Wang, X., Yu, K., Wu, S., Gu, J., Liu, Y., Dong, C., Qiao, Y., Loy, C.C.: Esrgan: Enhanced super-resolution generative adversarial networks. In: Leal-Taix{\'e}, L., Roth, S. (eds.) Computer Vision -- ECCV 2018 Workshops. pp. 63--79. Springer International Publishing, Cham (2019)

\bibitem{SADM}
Yoon, J.S., Zhang, C., Suk, H.I., Guo, J., Li, X.: Sadm: Sequence-aware diffusion model for longitudinal medical image generation. In: Information Processing in Medical Imaging (2022), \url{https://api.semanticscholar.org/CorpusID:254823541}

\end{thebibliography}

\end{document}